\documentclass[fleqn, usenatbib]{mnras}

\usepackage{newtxtext, newtxmath}
\usepackage[T1]{fontenc}
\usepackage{ae, aecompl}

\usepackage{graphicx} % Figures

\graphicspath{{figs/}}

\title[Radio eclipses of exoplanets]{Radio eclipses of exoplanets by the winds of their host stars}
\author[Kavanagh \& Vidotto]{R.~D.~Kavanagh$^{1}$\thanks{E-mail: \texttt{kavanar5@tcd.ie}}, 
A.~A.~Vidotto$^{1}$ \\
$^{1}$School of Physics, Trinity College Dublin, The University of Dublin, Dublin 2, Ireland \\
}

\date{Accepted XXX. Received YYY; in original form ZZZ}

\pubyear{2019}

\begin{document}
\label{firstpage}
\pagerange{\pageref{firstpage}--\pageref{lastpage}}
\maketitle

% ########################################################################
% Abstract
% ########################################################################

\begin{abstract}
The search for exoplanetary radio emission has resulted in zero conclusive detections to date. Various explanations for this have been proposed, from the observed frequency range, telescope sensitivity, to beaming of the emission. In a recent paper, we illustrated that exoplanets can orbit through the radio photosphere of the wind of the host star, a region that is optically thick at a specific frequency, for a large fraction of their orbits. As a result, radio emission originating from the planet could be absorbed or `eclipsed' by the wind of the host star. Here we investigate how the properties of the stellar wind and orbital parameters affect the fraction of the orbit where the planet is eclipsed by the stellar wind. We show that planets orbiting stars with low density winds are more favourable for detection in the radio. In terms of the orbital parameters, emission from transiting planets can escape the stellar wind easiest. We apply our model to the $\tau$~Boo planetary system, and show that observing the fraction of the planet's orbit where it is eclipsed by the wind of the host star could be used to constrain the properties of the stellar wind. However, our model developed would need to be used in conjunction with a separate method to disentangle the mass-loss rate and temperature of the stellar wind.
\end{abstract}

% Keywords

% ########################################################################
% Introduction
% ########################################################################

\section{Introduction}

Over the last few decades, the search for exoplanetary radio emission has been motivated by the desire for a new method of directly detecting exoplanets. In addition to this, if the mechanism driving the planetary radio emission is due to interactions with the planet's intrinsic magnetic field, detection of exoplanetary radio emission would allow us to assess the strength of the planet's magnetic field \citep{zarka01}.  This has consequences on the size of the planet's magnetosphere, which in turn can effect the lifetime of the planet's atmosphere. Whether this has a positive or negative effect is still in debate \citep{blackman18, carolan19, egan19}. 

Despite many attempts, there have been no conclusive detections of exoplanetary radio emission to date \citep[e.g.][]{lazio10, sirothia14, ogorman18}. Some possible explanations for this include the frequency range of the observations, the telescope sensitivity, the atmosphere of the planet \citep{weber18, daleyyates18}, and the emission not being beamed in the direction of the observer \citep{smith09}. Another possible explanation for this was made by \citet{kavanagh19}, who illustrated that exoplanets can orbit through the radio photosphere of the winds of their host stars, a region where a large fraction of radio emission can be absorbed through free-free processes. As the radio photosphere is not spherically symmetric, when the planet orbits further into the wind of the host star more of its emission can be absorbed.

There have been hints of a decrease in the flux density at radio frequencies from transiting exoplanets near secondary transit. \citet{smith09} observed this in the case of hot Jupiter HD189733b. Similarly, observations of HAT-P-11b system by \citet{lecavelier13} suggested a dip in the flux density near secondary transit. However, these results are inconclusive, and require follow-up observations with more sensitive instrumentation \citep{smith09, lecavelier13, sirothia14}. If confirmed, these dips in flux density near secondary transit could be due to the stellar wind of the host star absorbing the emission from the planet, as suggested by \citet{kavanagh19}.

A similar illustration of this phenomenon is the case of Black Widow pulsar systems \citep{roberts11}. These systems consist of a millisecond pulsar that is host to a low-mass companion with a mass of $\sim0.01~M_{\sun}$. As the companion nears primary transit of the pulsar host, radio emission originating from the pulsar is observed to disappear \citep[see][]{polzin19}. This is believed to be due to the wind of the low-mass companion star eclipsing the pulsar. For example, \citet{guillemot19} observed eclipses of radio emission at 1.4~GHz from the Black Widow pulsar PSR~J2055+3829. Due to the time variations of the eclipse, this has been attributed to the clumpy outflow of the companion star.

Radio eclipses have been observed in higher mass binaries as well. For example, \citet{dougherty05} observed a similar phenomenon in the massiv binary system WR140. From VLBA observations at 8.4~GHz they resolved that radio emission generated from the collision of the winds of the two star in the system disappears as the O-type star approaches periastron of the Wolf-Rayet star. This has been interpreted as the emission region entering the optically thick region of the wind of the Wolf-Rayet star.

In this paper, we present a model that predicts that radio emission from the planet becomes eclipsed by the wind of the host star as the planet progresses through its orbit. We illustrate how the properties of the wind of the host star and the planetary orbit affect the duration of the radio eclipse, and then apply our model to the $\tau$~Boo planetary system. We show how the model presented could be used to constrain properties such as the mass-loss rate and temperature of the stellar wind of the host star.

% ########################################################################
% Radio eclipse of the planet by the stellar wind
% ########################################################################

\section{Radio eclipse of the planet by the stellar wind}
\label{sec:radio eclipse}

To model the wind of the host star, we consider only forces due to the gravity of the star and the thermal pressure gradient of the wind \citep{parker58}:
\begin{equation}
\rho u \frac{d}{dr} u = - \frac{\rho G M_\star}{r^2} - \frac{d}{dr} p.
\label{eq:isothermal force}
\end{equation}
Here, $\rho$ is the mass density, $u$ is the velocity, $r$ is the distance, $G$ is the gravitational constant, $M_\star$ is the mass of the star, and $p$ is the thermal pressure. Assuming that the stellar wind is isothermal and mass is conserved, Equation~\ref{eq:isothermal force} can be re-arranged into the following form:
\begin{equation}
\frac{1}{u} \frac{d}{dr} u = \Big ( \frac{2 a^2}{r} - \frac{G M_\star}{r^2} \Big ) \Big / \Big ( u^2 - a^2 \Big ),
\label{eq:isothermal momentum}
\end{equation}
where $a$ is the isothermal sound speed:
\begin{equation}
a = \sqrt{ \frac{k_\text{B} T}{\mu m_\text{p}} }.
\end{equation}
Here, $k_\text{B}$ is Boltzmann's constant, $T$ is the wind temperature, and $\mu m_\text{p}$ is the mean mass per particle in the stellar wind, with $m_\text{p}$ being the proton mass. For a fully ionised hydrogen wind, we take $\mu = 1/2$.

We solve Equation~\ref{eq:isothermal momentum} enforcing a solution that passes through the critical point, where the numerator and denominator on the right-hand side go to zero simultaneously. This is to ensure a wind solution is obtained which monotonically increases in velocity outward \citep{parker58}. Such a critical point occurs at the sonic distance
\begin{equation}
r_\text{c} = \frac{G M_\star}{2 a^2},
\end{equation}
where the velocity of the wind at the critical point is $u = a$. Note that a physical solution is one where $r_\text{c}$ is greater than the stellar radius. With that, we obtain the wind velocity profile for a given stellar mass $M_\star$ and wind temperature $T$. The mass-loss rate of the stellar wind is
\begin{equation}
\dot{M} = 4 \pi r^2 \rho u.
\end{equation}
So, for a given stellar mass, wind temperature and mass-loss rate, we obtain a mass number density profile of the stellar wind:
\begin{equation}
n = \frac{\dot{M}}{4 \pi r^2 \mu m_\text{p} u} .
\label{eq:number density}
\end{equation}
One of the benefits of using a 1D isothermal stellar wind model is that it is much less time consuming than performing a 3D magnetohydrodynamic stellar wind simulation, such as those computed by \citet{kavanagh19}. A limitation of isothermal stellar wind models is that they neglect the presence of a magnetic field, which is important for determining the angular momentum-loss rates and therefore the rotational evolution of low-mass stars \citep{weber67, reville15, johnstone15a, ofionnagain19}. However, as we do not consider an evolving stellar wind here, we assume that an isothermal stellar wind model is sufficient for our purposes in this paper.

\begin{figure}
\centering
\includegraphics[width = \columnwidth]{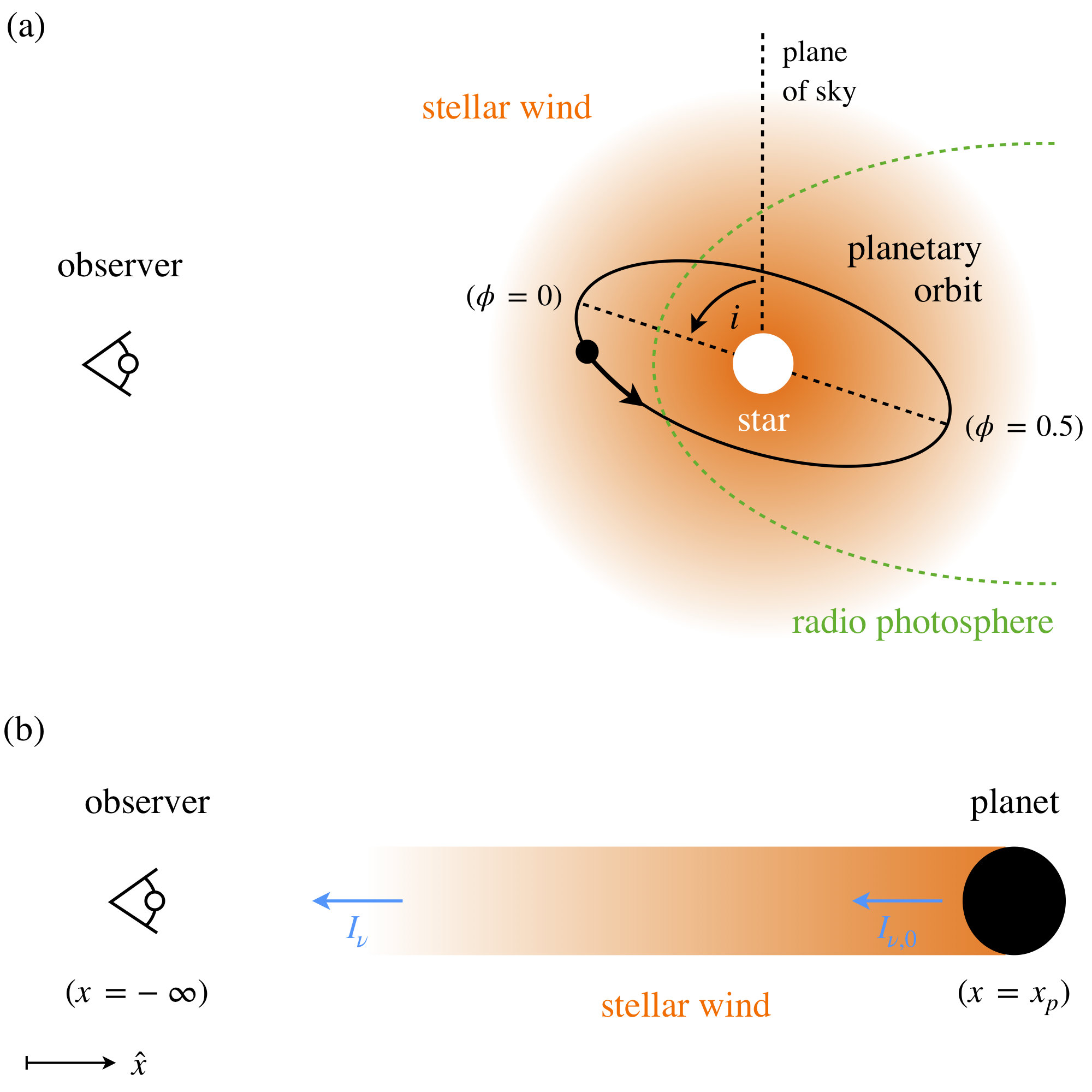}
\caption{\textit{Top}: Sketch illustrating a planet that orbits in a circle through the wind of its host star, inclined at an angle $i$ from the plane of the sky. Orbital phases of $\phi = 0$ and $\phi = 0.5$, corresponding to when the planet is nearest and furthest from the observer respectively, are marked along the orbit. The radio photosphere of the stellar wind is also shown, which the planet orbits through for a fraction of its orbit. \textit{Bottom}: The line of sight radio emission originating from the planet takes through the stellar wind towards the observer. The planet emits a specific intensity of $I_{\nu,0}$, and the observer sees an attenuated specific intensity of $I_\nu = I_{\nu,0} e^{-\tau_\nu}$.}
\label{fig:sketch}
\end{figure}

As the planet progresses further into its orbit, the amount of stellar wind material between the planet and observer increases. Since the wind of the host star can absorb low-frequency radio emission through free free processes \citep{panagia75}, a larger fraction of radio emission from the planet will be absorbed as it approaches an orbital phase of $\phi = 0.5$. We refer to a region of the stellar wind at a specific frequency as the `radio photosphere', wherein only a certain fraction (i.e.~50\%) of radio emission can escape through the wind. This is illustrated in the top panel of Figure~\ref{fig:sketch}. Note that there is one further complication: if the plasma frequency of the stellar wind at the planet's orbit 
\begin{equation}
\nu_p = 9 \times 10 ^ {-3} \sqrt{n_e} \ \text{MHz}
\end{equation}
is greater than the emitted frequency from the planet $\nu$, no emission will generate. Here, $n_e$ is the electron number density of the local stellar wind in cm$^{-3}$.

We do not model the emission mechanism of the planet here, but rather how such an emission, if existed, would be absorbed by the wind of the host star. We also assume that the emission from the planet is always beamed towards the observer. To determine how much planetary emission is absorbed by the stellar wind, we compute the optical depth at frequency $\nu$ at the planet's position $x_\text{p}$ along the line of sight in the stellar wind (see Figure~\ref{fig:sketch}):
\begin{equation}
\tau_\nu = \int_{-\infty}^{x_\text{p}} \alpha_\nu dx.
\label{eq:optical depth}
\end{equation}
Here, $\alpha_\nu$ is the free-free absorption coefficient \citep{cox00}:
\begin{equation}
\alpha_\nu = 3.692 \times 10 ^ 8 \Big ( 1 - e ^ {- h \nu / k_\text{B} T} \Big ) Z^2 g T^{-1/2} v^{-3} n_\text{e} n_\text{i} ,
\label{eq:abs coeff}
\end{equation}
where $h$ is Planck's constant, $Z$ is the ionisation state (+1 for ionised hydrogen), and $g$ is the Gaunt factor \citep{cox00}:
\begin{equation}
g = 10.6 + 1.9 \log_{10} (T) - 1.26 \log_{10} (Z \nu) .
\end{equation}
In Equation~\ref{eq:abs coeff}, $n_\text{e}$ and $n_\text{i}$ are the electron and ion number densities respectively. Since we treat the wind as being composed of fully ionised hydrogen, the total number density is $n = n_\text{e} + n_\text{i} = 2 n_\text{e}$ ($n_\text{e} = n_\text{i}$). The ion number density is the proton number density in this case.

The observer receives a specific intensity $I_\nu$ at frequency $\nu$ from the planet that has been attenuated by the stellar wind \citep{rybicki86}:
\begin{equation}
I_\nu = I_{\nu,0} e ^ {-\tau_\nu} ,
\label{eq:specific intensity}
\end{equation}
where $I_{\nu,0}$ is the emitted specific intensity by the planet. This is illustrated in the bottom panel of Figure~\ref{fig:sketch}. The flux density received by the observer is then:
\begin{equation}
F_\nu = \int I_\nu d\Omega = \frac{1}{d^2} \int I_\nu dA,
\label{eq:flux int}
\end{equation}
where $d\Omega$ is the element of solid angle of the emitting region, $d$ is the distance to the system, and $dA$ is the area element of the emitting region. Assuming that the specific intensity $I_\nu$ is constant through $dA$, then from Equations~\ref{eq:specific intensity} and \ref{eq:flux int}:
\begin{equation}
F_\nu \propto e^{- \tau_\nu} .
\end{equation}
So, if $F_{\nu, 0}$ is the flux density one would observe in the absence of a stellar wind, the observed flux density from the planet is:
\begin{equation}
F_\nu = F_{\nu, 0} e^{- \tau_\nu} .
\label{eq:flux density}
\end{equation}
The term $e^{- \tau_\nu}$ therefore gives the fraction of the flux density that is transmitted from the planet to the observer. When computing the optical depth at each point in the planet's orbit with Equation~\ref{eq:optical depth}, we extend the stellar wind out from the planet to a distance of 0.5~au in the direction of the observer ($-x$ direction). This distance is sufficient for the optical depth at the planet's position to converge. 

\begin{figure*}
\centering
\includegraphics[width = 0.495\textwidth]{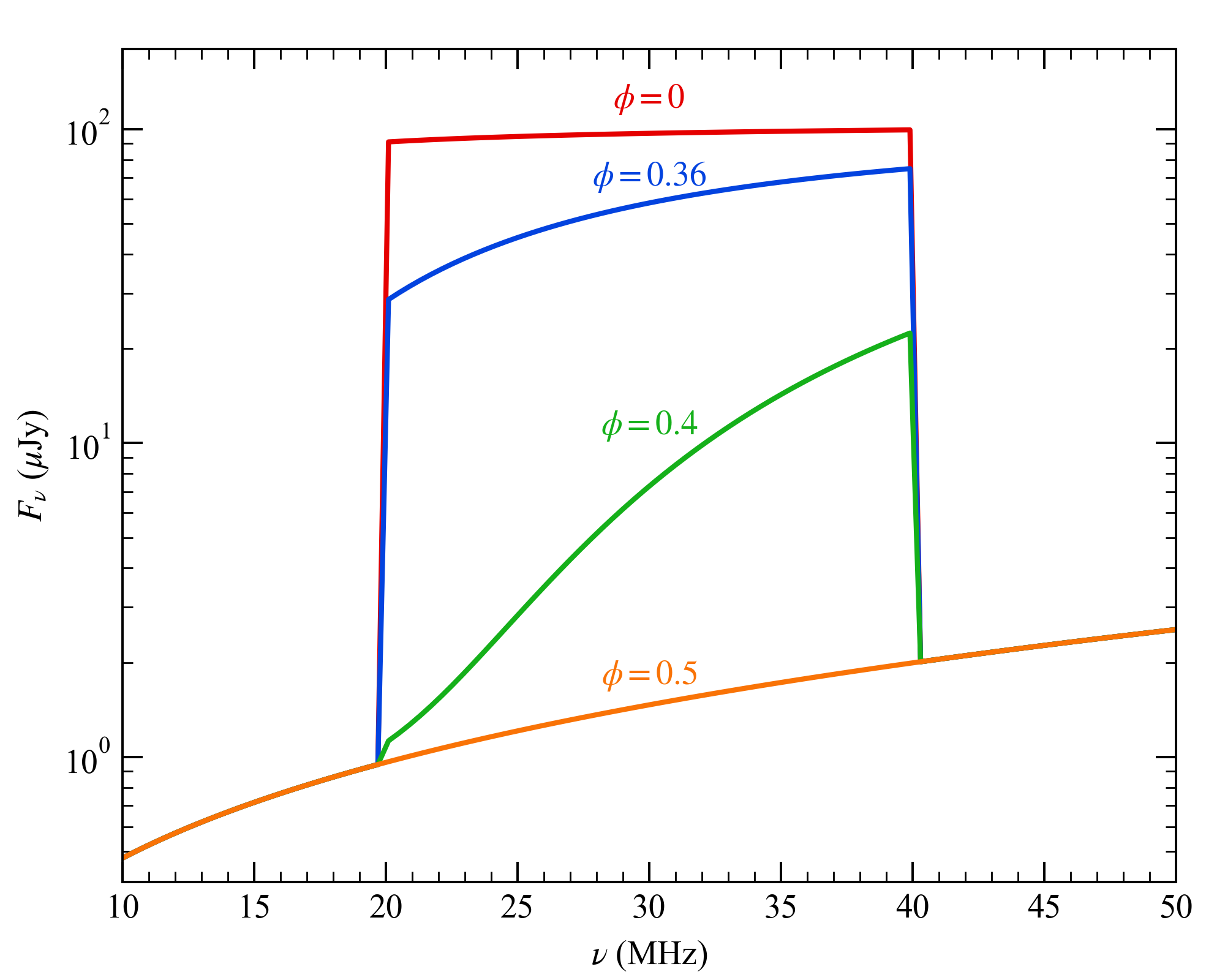}
\includegraphics[width = 0.495\textwidth]{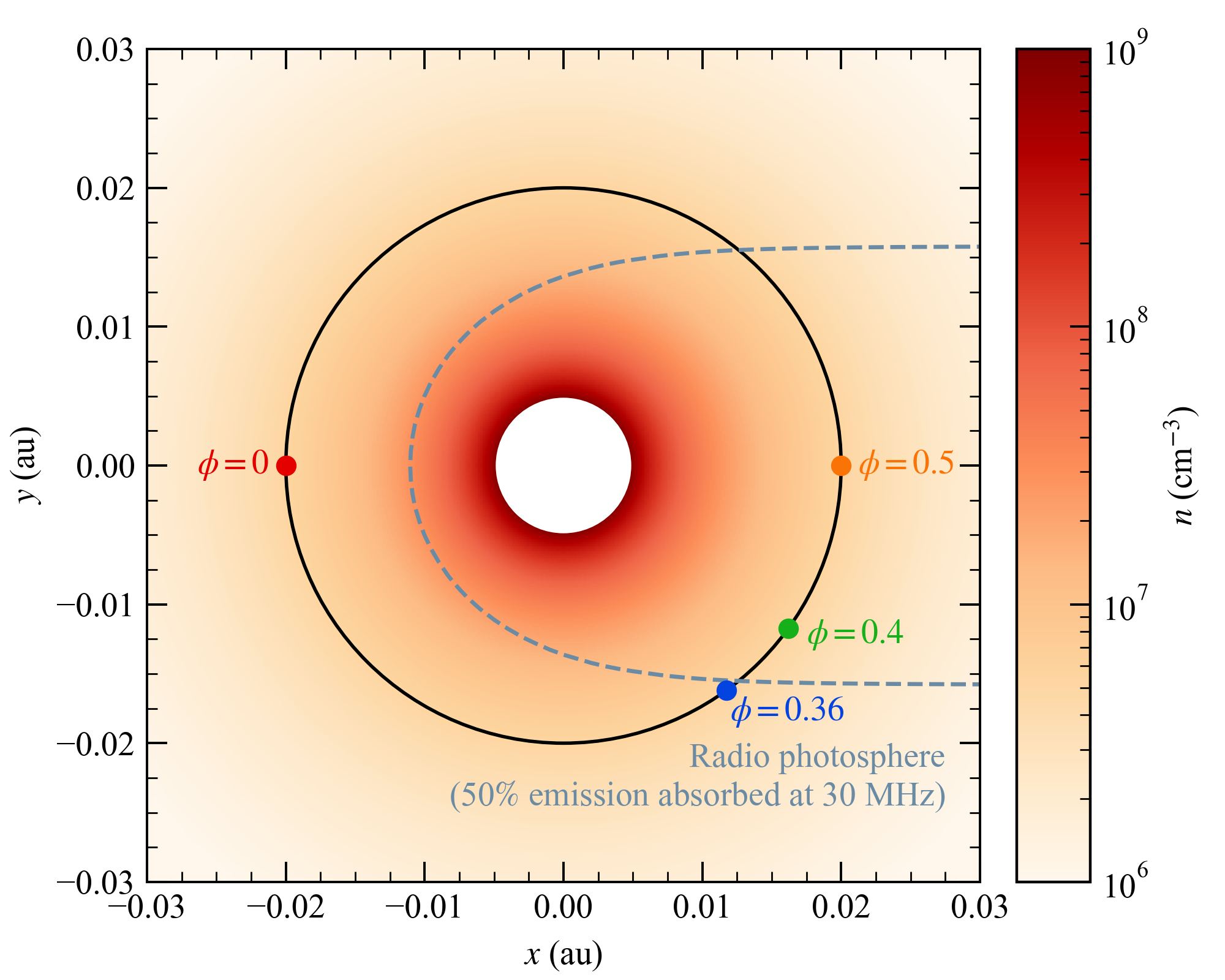}
\caption{\textit{Left:} The combined spectrum of free-free emission from the stellar wind and emission from the planet at different orbital phases. The system is placed at 1~pc for this illustration, with the planet emitting an arbitrary unattenuated flux density of 100~$\mu$Jy. \textit{Right:} Number density profile of the stellar wind, and the positions of the planet corresponding to the spectra shown in the left panel. The dashed line shows the radio photosphere where 50\% of emission is absorbed at 30~MHz by the stellar wind. The observer looks towards the system from $x=-1$~pc along the $x$-direction.}
\label{fig:spectrum}
\end{figure*}

Figure~\ref{fig:spectrum} illustrates how emission from a transiting planet orbiting its host star disappears as it approaches secondary transit. In this example, the planet orbits a star with a mass of $1~M_{\sun}$ and radius of $1~R_{\sun}$ at 0.02~au. The wind of the host star has a mass-loss rate of $100~\dot{M}_{\sun}$, where $\dot{M}_{\sun}$ is the solar wind mass-loss rate ($2\times 10^{-14} \ M_{\sun}\ \text{yr}^{-1}$), and a temperature of 2~MK. The left panel shows the combined thermal spectrum of the wind of the star with that of a planet at different orbital phases. We assume that the planet emits an unattenuated flux density of $F_{\nu,0} = 100$~$\mu$Jy in the region of 20-40~MHz. The value of $F_{\nu,0}$ chosen here is purely for illustrative purposes. The thermal spectrum of the stellar wind is computed as per the method laid out in \citet{ofionnagain19} and \citet{kavanagh19}. We compute the spectrum placing the system at 1~pc from the observer.

The right panel of Figure~\ref{fig:spectrum} shows the number density profile of the stellar wind which the planet orbits through. The orbital phases we compute the spectrum of the planet for in the left panel are marked along the orbit. As the planet progresses through its orbit towards secondary transit ($\phi \gtrsim 0.36$), the flux received from the planet begins to disappear. Past a certain point the wind of the host star eclipses the planet. Towards higher frequencies however, this effect is less pronounced. This was shown by \citet{kavanagh19}, in that the planet is easier to detect both near primary transit of the host star and at higher emitted frequencies.

% Sketch

% Spectrum of wind + planet

% ########################################################################
% Varying stellar wind properties
% ########################################################################

\section{How do the stellar wind properties affect how much planetary emission escapes?}
\label{sec:sw properties}

\begin{figure}
\centering
\includegraphics[width = \columnwidth]{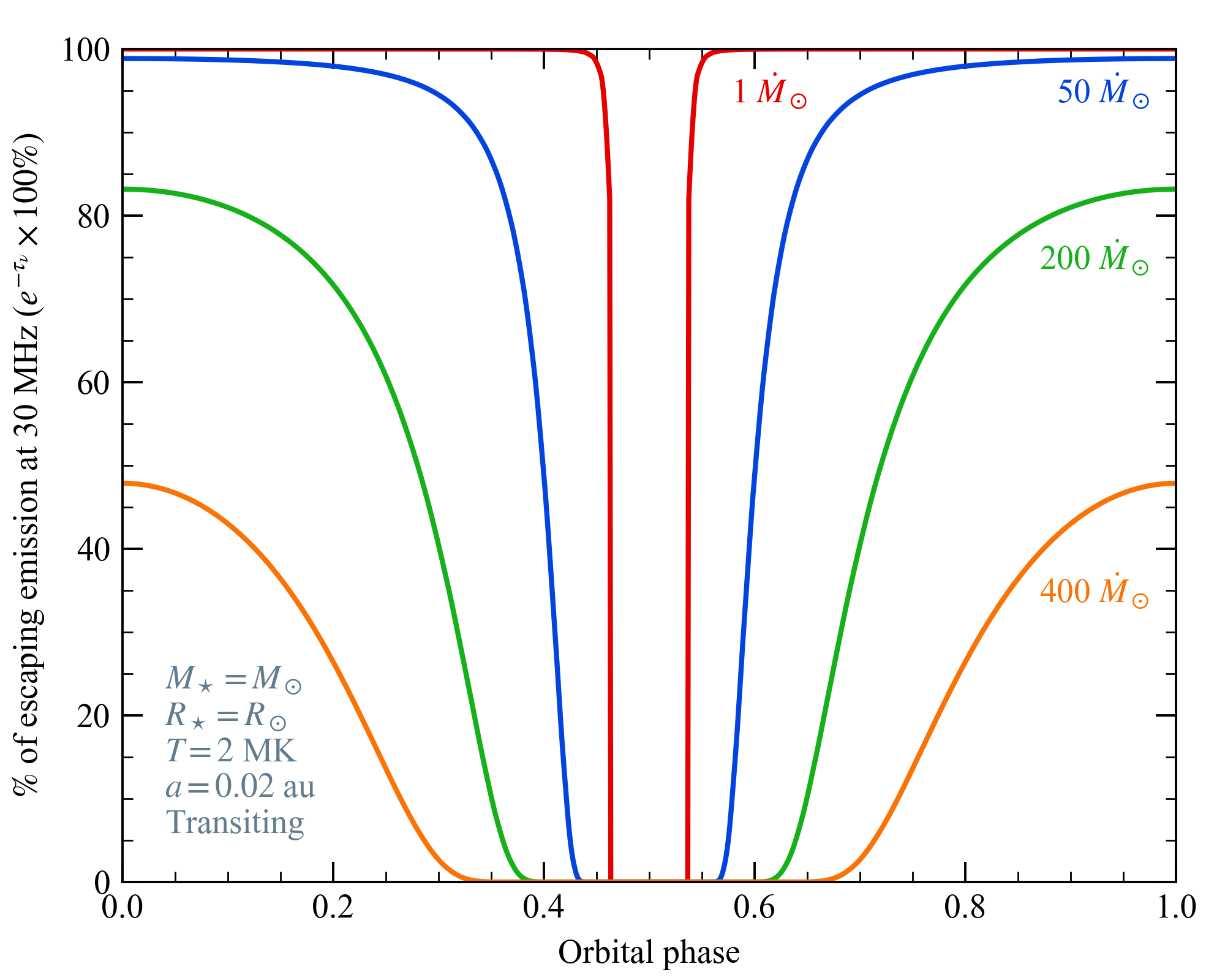}
\caption{The percentage of planetary emission at 30~MHz ($e^{-\tau_\nu}\times100\%$) that escapes through the stellar wind as a function of orbital phase for different orbital inclinations. The stellar, wind, and orbital parameters are listed in the bottom left corner. When $\sim0\%$ of the emission escapes, the planet is considered to be in `radio eclipse'. Note that the fraction of the orbit during radio eclipse is larger for winds of higher mass-loss rates.}
\label{fig:vary mdot}
\end{figure}

Here, we investigate how the properties of the stellar wind itself affect how much planetary emission escapes through the wind. Figure~\ref{fig:vary mdot} shows how varying the mass-loss rate of the wind of the host star affects the percentage of escaping emission at 30~MHz as the planet progresses through its orbit. As the planet approaches secondary transit of the star (orbital phase of 0.5), the planetary emission becomes eclipsed by the stellar wind. This effect is more pronounced for higher stellar wind mass-loss rates, which in turn are denser for a fixed temperature. Note that at 30~MHz, a low mass-loss rate stellar wind does not attenuate the planetary emission at primary transit ($\phi = 0$). However at higher mass-loss rates, even when the planet transits part of its emission is attenuated and the flux density would not be observed at this frequency.

\begin{figure}
\centering
\includegraphics[width = \columnwidth]{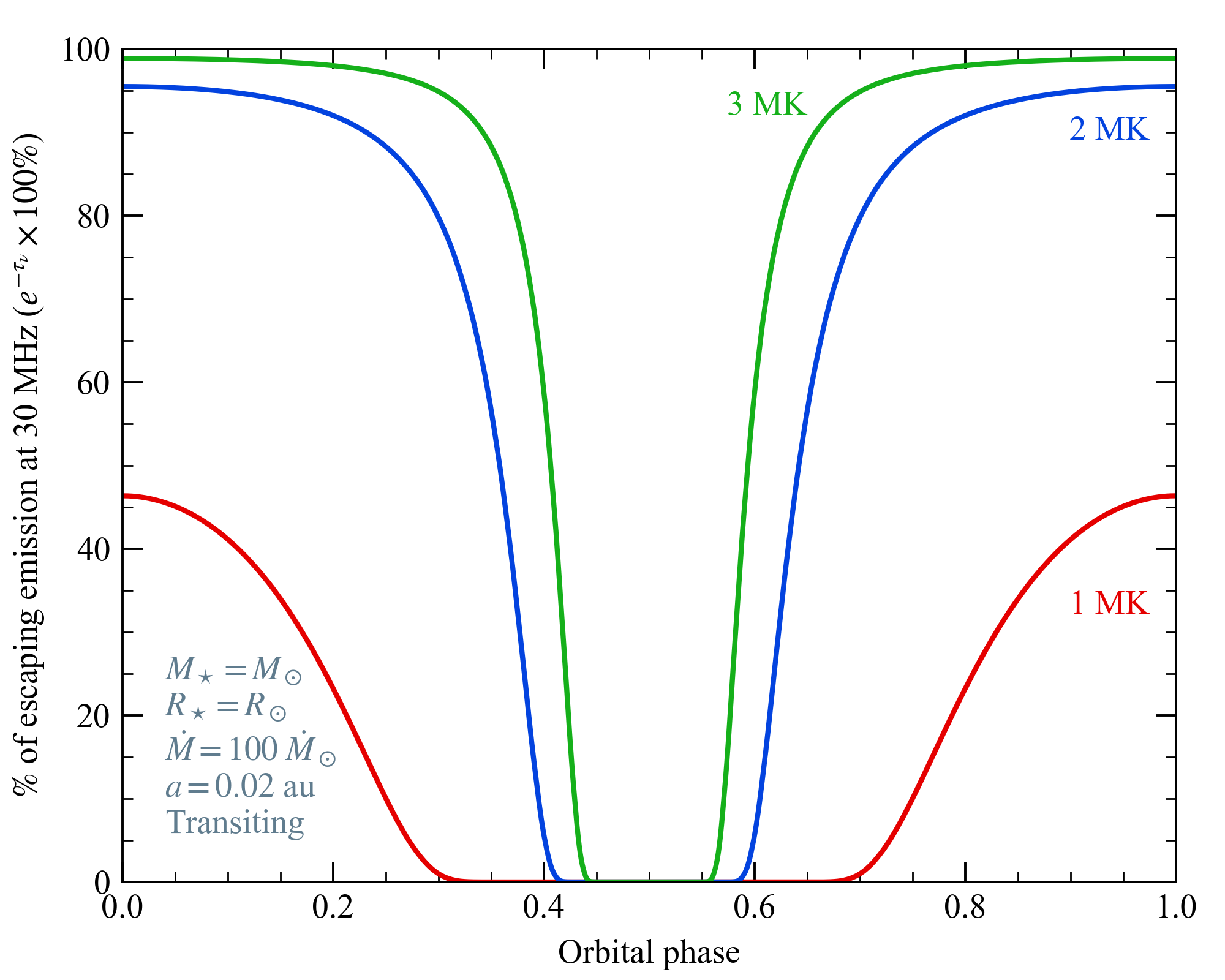}
\caption{Same as Figure~\ref{fig:vary mdot}, but for different stellar wind temperatures.}
\label{fig:vary temp}
\end{figure}

In Figure~\ref{fig:vary temp} we show varying how the temperature of the stellar wind, for a fixed mass-loss rate, affects the amount of escaping emission from the planet. We see that detection of exoplanetary radio emission is favoured for high temperature winds, with the eclipse occurring for a smaller fraction of the planet's orbit. As the wind is accelerated much faster in at higher temperatures, the density of the wind drops off much faster (following from Equation~\ref{eq:number density}). As a result, hotter winds tend to be less dense. However, the mass-loss rates of low-mass stars are expected to increase with surface X-ray flux \citep[see][]{wood18}, which in turn is correlated with a hot corona \citep{johnstone15b}. So while the most favourable detection scenario would be a planet that orbits a star with a hot wind with a low mass-loss rate, it is more likely that if the wind is very hot, the mass-loss rate is high. In general, planets orbiting stars with low density winds are more favourable for detection.

% Varying mass-loss rate

% Varying temperature

% ########################################################################
% Varying orbital characteristics
% ########################################################################

\section{How does the geometry of the planetary orbit affect how much emission escapes?}
\label{sec:orbital geometry}

\begin{figure}
\centering
\includegraphics[width = \columnwidth]{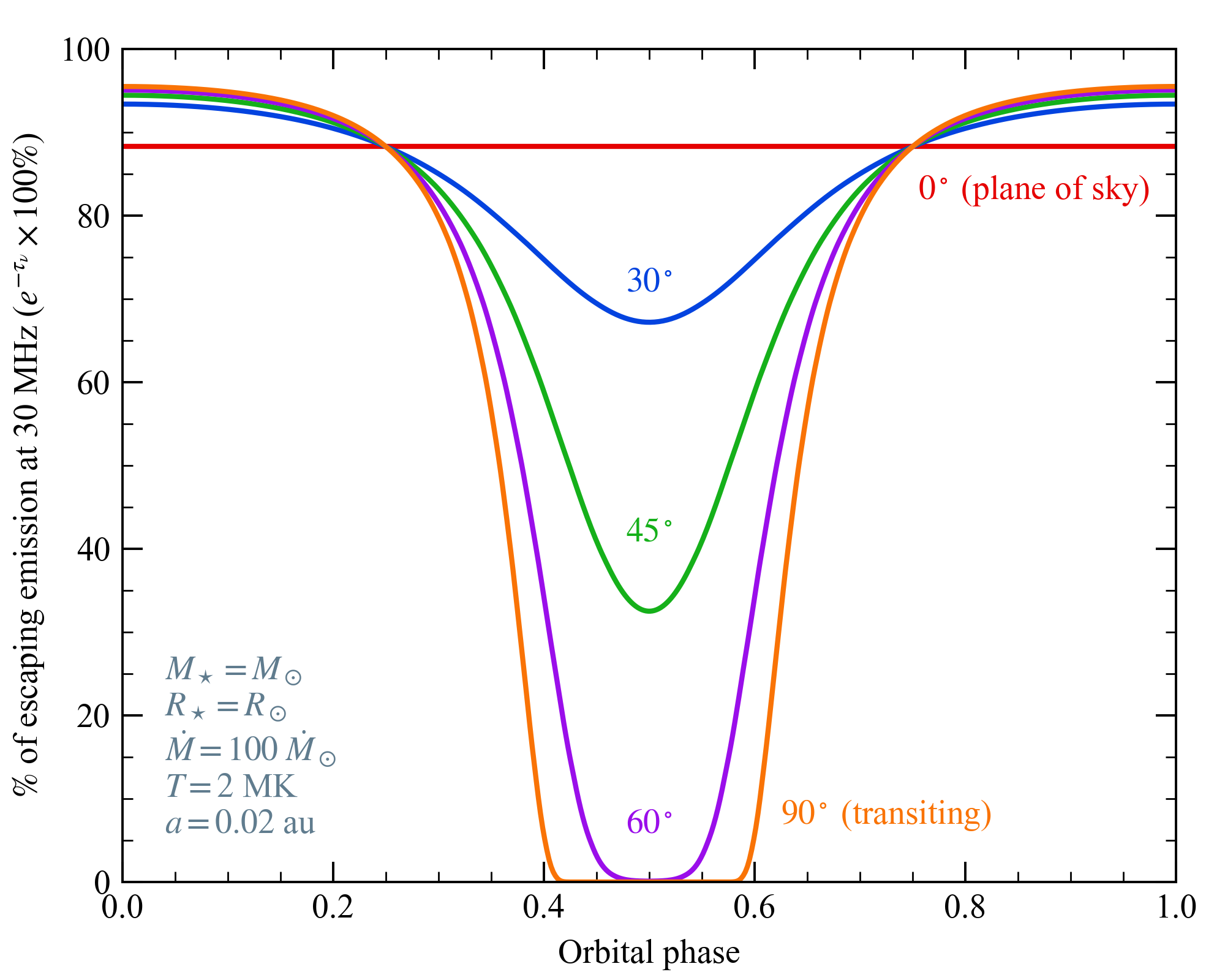}
\caption{Same as Figure~\ref{fig:vary mdot}, but for different orbital inclinations. Transiting planets have the most favourable conditions for detection, but also have the largest modulation of their emitted flux density.}
\label{fig:vary inc}
\end{figure}

The amount of planetary emission that escapes the wind of the host star also depends on the path it takes as the planet progresses through its orbit. Here, we investigate how the geometry of the orbit affects the duration of the radio eclipses. In Figure~\ref{fig:vary inc} we show the effects of varying the orbital inclination of the planet. We find that a transiting planet is the most easily detectable, specifically near primary transit ($\phi = 0$). However, in this configuration the planetary emission also is the most attenuated, near secondary transit ($\phi = 0.5$), with the largest amplitude of the eclipse modulation.  

The transmitted flux density on the other hand for a planet in a plane of sky orbital configuration is constant, assuming the unattenuated flux density $F_{\nu,0}$ is constant. As a result, a planet orbiting in the plane of the sky will either always be detectable or never be detectable, depending on the instrument sensitivity and the amount of flux that escapes the stellar wind. Detection is also in theory always possible for planets orbiting at inclinations $0 < i < \cos^{-1} (a / R_\star)$, as the planet does not pass behind the stellar disk.

\begin{figure}
\centering
\includegraphics[width = \columnwidth]{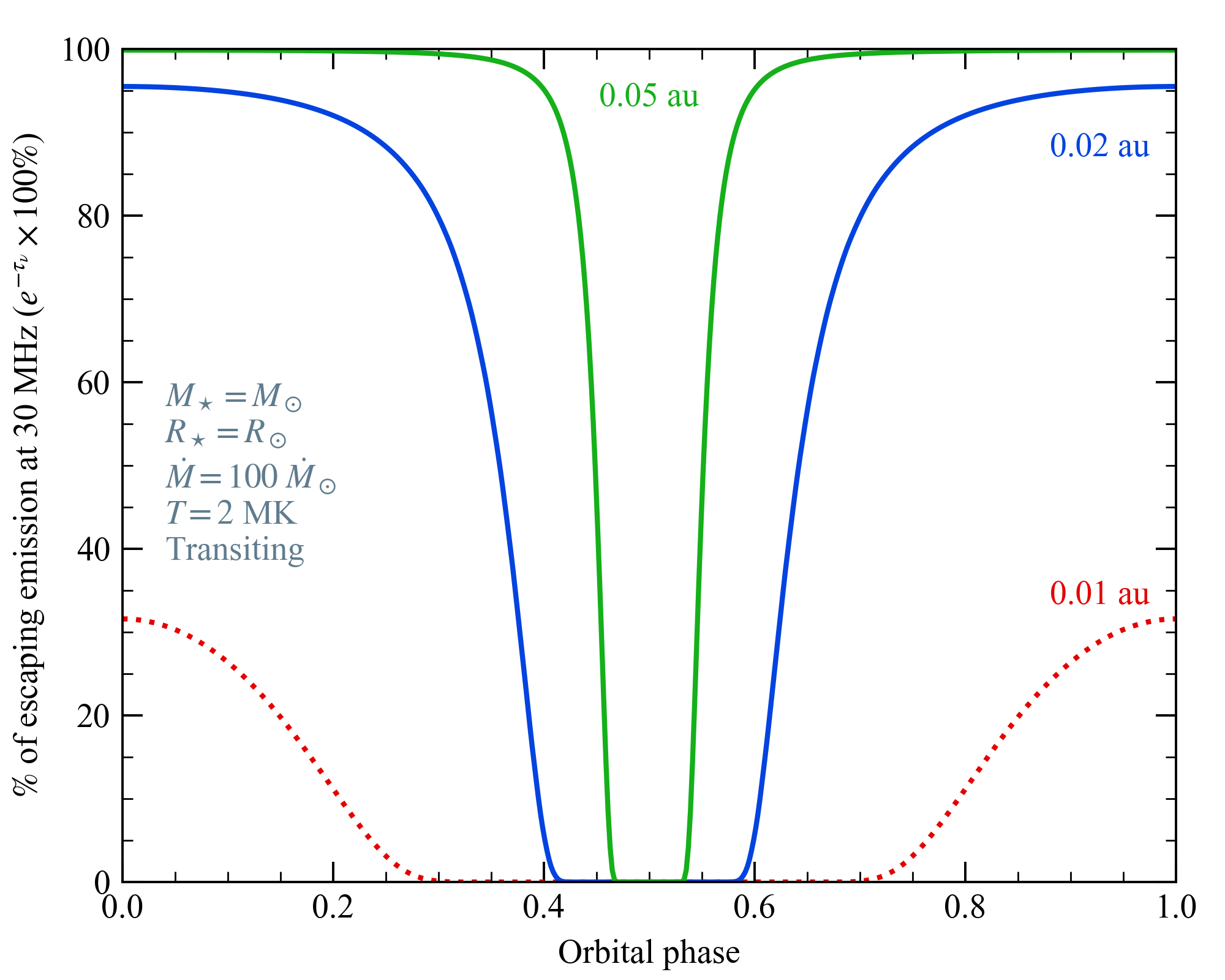}
\caption{Same as Figure~\ref{fig:vary mdot}, but for different orbital distances. At 0.01~au, the plasma frequency of the stellar wind at the planet's is greater than 30~MHz and as a result no emission can generate. This is indicated by the dashed line.}
\label{fig:vary dist}
\end{figure}

Figure~\ref{fig:vary dist} show the effects of varying the orbital distance of a transiting planet on the amount of escaping emission. We see that the closer in to the star the planet orbits, the lower the amount of emission escapes as the density of the stellar wind is much larger close in. In addition to this, if the stellar wind plasma frequency is too high the emission will not generate. This is the case for a planet orbiting at 0.01~au for the parameters presented in Figure~\ref{fig:vary dist}. So, planets orbiting further out are more easily detectable. However, if the stellar wind power dissipated onto the planetary magnetosphere powers the planetary radio emission, planets further out are likely to emit much lower flux densities than those closer in \citep{zarka01}. 

% ########################################################################
% Application of model to Tau Boo b
% ########################################################################

\section{Constraining the stellar wind properties of the hot Jupiter host $\tau$ Boo}
\label{sec:tauboo}

We now focus on the $\tau$~Boo planetary system, which lies 15.6~pc away from Earth. The star is host to the supermassive hot Jupiter $\tau$~Boo~b which orbits at just 0.046~au, and as a result the planet is expected to be a source of strong radio emission \citep[see][]{weber18}. As the planet does not transit its host star ($i = 45^\circ$), in theory it could be detectable for the entirety of its orbit if it emits a high enough flux density, following the results shown in Figure~\ref{fig:vary inc}. The mass and radius of the star are 1.30~$M_{\sun}$ and 1.33~$R_{\sun}$ respectively. All parameters listed above are taken from \texttt{exoplanet.eu}.

As illustrated in Figures~\ref{fig:vary mdot} and \ref{fig:vary temp}, the amount of escaping emission from the planet depends on the mass-loss rate and temperature of the wind of the host star, both of which are unknowns in the case of $\tau$~Boo. Another two unknown quantities of the system are the emitted flux density and frequency from the planet (if it is indeed a source of exoplanetary radio emission). \citet{turner19} showed that emission from $\tau$~Boo~b would be detectable with LOFAR, if it emits a flux density that is $10^5$ times larger than that emitted by Jupiter. During active periods, Jupiter has an observed flux density $10^7$~Jy at a distance of 1~au \citep{zarka04}. Therefore, at a distance of 15.6~pc, we assume an unattenuated flux density for $\tau$~Boo~b of $F_{\nu,0} = 96.6$~mJy.

If we also assume that the planet emits at a frequency of 30~MHz, we can determine for what fraction of the planet's orbit it would be detectable with LOFAR. At 30~MHz LOFAR has a sensitivity of $F_\nu^\text{LOFAR}\sim7$~mJy for a 1~hour integration time \citep[see Figure~1 of][]{griessmeier11}. The planet is then detectable when $F_\nu > F_\nu^\text{LOFAR}$ or when $e^{-\tau_\nu} > 0.0725$ following Equation~\ref{eq:flux density}.

\begin{figure}
\centering
\includegraphics[width = \columnwidth]{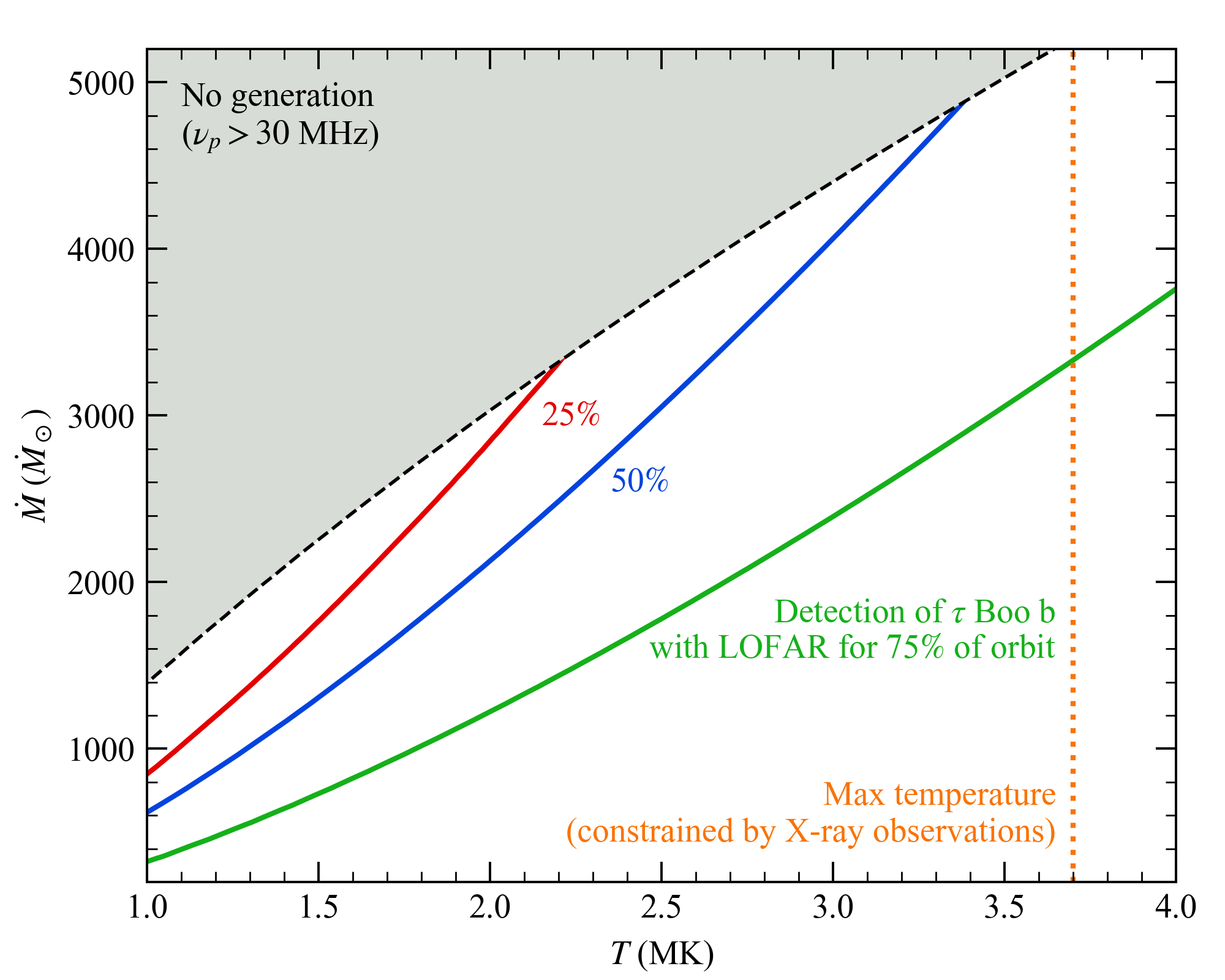}
\caption{The combined stellar wind mass-loss rate and temperature that results in detection of radio emission from $\tau$~Boo~b at 30~MHz with LOFAR, for 25\%, 50\%, and 75\% of its orbit. The shaded area illustrates where the plasma frequency of the stellar wind at the planet's orbit is too high for generation of radio emission at 30~MHz. The vertical dotted line marks the maximum temperature of the stellar wind, which is constrained by X-ray observations of the host star (see relevant text).}
\label{fig:tauboo}
\end{figure}

Figure~\ref{fig:tauboo} shows the combination of stellar wind mass-loss rate and temperature that results in detection of emission from $\tau$~Boo~b at 30~MHz with LOFAR for 25\%, 50\%, and 75\% of its orbit. As can be seen, if emission was detected from the system at 30~MHz for, say, 50\% of the planet's orbit, there is a degeneracy between the stellar wind mass-loss rate and temperature. So while observing radio emission alone from the planet for a certain fraction of its orbit does not provide enough information to derive the stellar wind properties, it could nevertheless be used in conjunction with the radiometric Bode's law to provide additional constraints \citep[see][]{vidotto17}. In Figure~\ref{fig:tauboo} we also show that the plasma frequency of the stellar wind at the planet's orbit and the maximum temperature of the stellar wind can constrain the parameters of the stellar wind. We derive a maximum coronal temperature (and therefore the wind) for $\tau$~Boo of 3.7~MK from \citet{johnstone15b}, using the maximum observed X-ray luminosity for the host star of $8\times10^{28}$~erg~s$^{-1}$ \citep{mittag17}.

The mass-loss rates of the winds of low-mass stars have proved to be very difficult to measure, and has only been possible in a handful of cases. One such method relies on observations of the absorption of Ly-$\alpha$ from the star due to the build-up of neutral hydrogen at the astrosphere \citep{wood04}. \citet{vidotto17b} also illustrated how stellar wind mass-loss rates could be derived based on observations of Ly-$\alpha$ absorption due to the presence of an extended planetary atmosphere. Both of these methods however require Hubble Space Telescope observations, and therefore cannot be carried out for a large number of targets. More recently, \citet{jardine19}  showed how the mass-loss rates of low-mass stars could be derived based on the presence of prominences from H$\alpha$ observations. This method however is only possible for very fast rotators. Mass-loss rates from low-mass stars could also be determined from observations of thermal free-free radio emission from the stellar wind itself \citep{gudel02}, however current radio telescopes are not sensitive enough to detect this emission \citep{ofionnagain19}. Our method presented here could provide a new method of constraining the mass-loss rates of low-mass stars from radio observations with current telescopes (i.e., LOFAR), if combined with another method such as the radiometric Bode's law.

One of the main issues with the application of our model in this section is that in order to determine when a planet would be detectable with a given radio telescopes, we require knowledge of the unattenuated flux density emitted by the planet. In addition to this, it is likely that the emitted flux density varies over the orbit. This has been shown to be the case if the emission mechanism powering the planetary radio emission is the interaction between the stellar wind and planetary magnetosphere, as the wind of the host star is not uniform \citep[see Figure 4 of][]{kavanagh19, nicholson16}. For instance, Jupiter is observed to experience active periods where the flux density emitted increases \citep{zarka04}.

% ########################################################################
% Conclusions
% ########################################################################

\section{Conclusions}

In this work, we have presented how the properties of the stellar wind and the orbital characteristics of a planet affect the fraction of the planet's orbit where its radio emission is eclipsed by the wind of the host star. We have illustrated that detection of planetary radio emission is favoured for host stars with low density winds. The ideal case for this would be a star with a hot wind and low-mass loss rate, however low-mass stars with hot winds generally have high mass-loss rates.

In terms of orbital parameters, emission can most easily escape the wind of the host star for transiting planets, specifically in the region near primary transit. However, in this orbital configuration, the emission is also most attenuated near secondary transit. Therefore, transiting exoplanets offer the most extreme case in which radiation escapes the most ($\phi = 0$) or is most attenuated ($\phi = 0.5$). Planets with orbital inclinations in the range $0 < i < \cos^{-1} (a / R_\star)$ can in theory always be detected, depending on the emitted flux density and density of the stellar wind, as they do not pass behind the stellar disk. We also have illustrated that planets orbiting closer to their host stars are less likely to be detected, as the stellar wind is very dense close in. However, if the planet orbits far out from its host star, the mechanism powering the radio emission of the planet may be too weak \citep[see][]{zarka01}.

We applied our model to the $\tau$~Boo planetary system, and illustrated that if radio emission is detected from the planet for a certain percentage of the orbit with LOFAR, we can constrain the wind properties of the host star. Observations such as this are not sufficient to disentangle the mass-loss rate and temperature of the wind. In combination with another method however such as the radiometric Bode's law could help to further constrain the stellar wind properties. The plasma frequency of the stellar wind as well as X-ray observations of the host star could also aid constraining the wind properties.

% ########################################################################
% Acknowledgements
% ########################################################################

\section*{Acknowledgements}

RDK acknowledges funding received from the Irish Research Council through the Government of Ireland Postgraduate Scholarship Programme. AAV acknowledges funding from the Irish Research Council Consolidator Laureate Award 2018 and the European Research Council (ERC) under the European Union's Horizon 2020 research and innovation programme (grant agreement No 817540, ASTROFLOW).

% ########################################################################
% Bibliography
% ########################################################################

\bibliographystyle{mnras}
\bibliography{bibliography}

\begin{thebibliography}{}
\makeatletter
\relax
\def\mn@urlcharsother{\let\do\@makeother \do\$\do\&\do\#\do\^\do\_\do\%\do\~}
\def\mn@doi{\begingroup\mn@urlcharsother \@ifnextchar [ {\mn@doi@}
  {\mn@doi@[]}}
\def\mn@doi@[#1]#2{\def\@tempa{#1}\ifx\@tempa\@empty \href
  {http://dx.doi.org/#2} {doi:#2}\else \href {http://dx.doi.org/#2} {#1}\fi
  \endgroup}
\def\mn@eprint#1#2{\mn@eprint@#1:#2::\@nil}
\def\mn@eprint@arXiv#1{\href {http://arxiv.org/abs/#1} {{\tt arXiv:#1}}}
\def\mn@eprint@dblp#1{\href {http://dblp.uni-trier.de/rec/bibtex/#1.xml}
  {dblp:#1}}
\def\mn@eprint@#1:#2:#3:#4\@nil{\def\@tempa {#1}\def\@tempb {#2}\def\@tempc
  {#3}\ifx \@tempc \@empty \let \@tempc \@tempb \let \@tempb \@tempa \fi \ifx
  \@tempb \@empty \def\@tempb {arXiv}\fi \@ifundefined
  {mn@eprint@\@tempb}{\@tempb:\@tempc}{\expandafter \expandafter \csname
  mn@eprint@\@tempb\endcsname \expandafter{\@tempc}}}

\bibitem[\protect\citeauthoryear{{Blackman} \& {Tarduno}}{{Blackman} \&
  {Tarduno}}{2018}]{blackman18}
{Blackman} E.~G.,  {Tarduno} J.~A.,  2018, \mn@doi [\mnras]
  {10.1093/mnras/sty2640}, \href
  {https://ui.adsabs.harvard.edu/abs/2018MNRAS.481.5146B} {481, 5146}

\bibitem[\protect\citeauthoryear{{Carolan}, {Vidotto}, {Loesch}  \&
  {Coogan}}{{Carolan} et~al.}{2019}]{carolan19}
{Carolan} S.,  {Vidotto} A.~A.,  {Loesch} C.,   {Coogan} P.,  2019, \mn@doi
  [\mnras] {10.1093/mnras/stz2422}, \href
  {https://ui.adsabs.harvard.edu/abs/2019MNRAS.489.5784C} {489, 5784}

\bibitem[\protect\citeauthoryear{{Cox}}{{Cox}}{2000}]{cox00}
{Cox} A.~N.,  2000, {Allen's Astrophysical Quantities}

\bibitem[\protect\citeauthoryear{{Daley-Yates} \& {Stevens}}{{Daley-Yates} \&
  {Stevens}}{2018}]{daleyyates18}
{Daley-Yates} S.,  {Stevens} I.~R.,  2018, \mn@doi [\mnras]
  {10.1093/mnras/sty1652}, \href
  {https://ui.adsabs.harvard.edu/abs/2018MNRAS.479.1194D} {479, 1194}

\bibitem[\protect\citeauthoryear{{Dougherty}, {Beasley}, {Claussen}, {Zauderer}
   \& {Bolingbroke}}{{Dougherty} et~al.}{2005}]{dougherty05}
{Dougherty} S.~M.,  {Beasley} A.~J.,  {Claussen} M.~J.,  {Zauderer} B.~A.,
  {Bolingbroke} N.~J.,  2005, \mn@doi [\apj] {10.1086/428494}, \href
  {https://ui.adsabs.harvard.edu/abs/2005ApJ...623..447D} {623, 447}

\bibitem[\protect\citeauthoryear{{Egan}, {Jarvinen}, {Ma}  \& {Brain}}{{Egan}
  et~al.}{2019}]{egan19}
{Egan} H.,  {Jarvinen} R.,  {Ma} Y.,   {Brain} D.,  2019, \mn@doi [\mnras]
  {10.1093/mnras/stz1819}, \href
  {https://ui.adsabs.harvard.edu/abs/2019MNRAS.488.2108E} {488, 2108}

\bibitem[\protect\citeauthoryear{{Grie{\ss}meier}, {Zarka}  \&
  {Girard}}{{Grie{\ss}meier} et~al.}{2011}]{griessmeier11}
{Grie{\ss}meier} J.~M.,  {Zarka} P.,   {Girard} J.~N.,  2011, \mn@doi [Radio
  Science] {10.1029/2011RS004752}, \href
  {https://ui.adsabs.harvard.edu/abs/2011RaSc...46.0F09G} {46, RS0F09}

\bibitem[\protect\citeauthoryear{{G{\"u}del}}{{G{\"u}del}}{2002}]{gudel02}
{G{\"u}del} M.,  2002, \mn@doi [\araa]
  {10.1146/annurev.astro.40.060401.093806}, \href
  {https://ui.adsabs.harvard.edu/abs/2002ARA&A..40..217G} {40, 217}

\bibitem[\protect\citeauthoryear{{Guillemot}, {Octau}, {Cognard}, {Desvignes},
  {Freire}, {Smith}, {Theureau}  \& {Burnett}}{{Guillemot}
  et~al.}{2019}]{guillemot19}
{Guillemot} L.,  {Octau} F.,  {Cognard} I.,  {Desvignes} G.,  {Freire}
  P.~C.~C.,  {Smith} D.~A.,  {Theureau} G.,   {Burnett} T.~H.,  2019, \mn@doi
  [\aap] {10.1051/0004-6361/201936015}, \href
  {https://ui.adsabs.harvard.edu/abs/2019A&A...629A..92G} {629, A92}

\bibitem[\protect\citeauthoryear{{Jardine} \& {Collier Cameron}}{{Jardine} \&
  {Collier Cameron}}{2019}]{jardine19}
{Jardine} M.,  {Collier Cameron} A.,  2019, \mn@doi [\mnras]
  {10.1093/mnras/sty2872}, \href
  {http://adsabs.harvard.edu/abs/2019MNRAS.482.2853J} {482, 2853}

\bibitem[\protect\citeauthoryear{{Johnstone} \& {G{\"u}del}}{{Johnstone} \&
  {G{\"u}del}}{2015}]{johnstone15b}
{Johnstone} C.~P.,  {G{\"u}del} M.,  2015, \mn@doi [\aap]
  {10.1051/0004-6361/201425283}, \href
  {https://ui.adsabs.harvard.edu/abs/2015A&A...578A.129J} {578, A129}

\bibitem[\protect\citeauthoryear{{Johnstone}, {G{\"u}del}, {Brott}  \&
  {L{\"u}ftinger}}{{Johnstone} et~al.}{2015}]{johnstone15a}
{Johnstone} C.~P.,  {G{\"u}del} M.,  {Brott} I.,   {L{\"u}ftinger} T.,  2015,
  \mn@doi [\aap] {10.1051/0004-6361/201425301}, \href
  {https://ui.adsabs.harvard.edu/abs/2015A&A...577A..28J} {577, A28}

\bibitem[\protect\citeauthoryear{{Kavanagh} et~al.,}{{Kavanagh}
  et~al.}{2019}]{kavanagh19}
{Kavanagh} R.~D.,  et~al., 2019, \mn@doi [\mnras] {10.1093/mnras/stz655}, \href
  {https://ui.adsabs.harvard.edu/abs/2019MNRAS.485.4529K} {485, 4529}

\bibitem[\protect\citeauthoryear{{Lazio}, {Shankland}, {Farrell}  \&
  {Blank}}{{Lazio} et~al.}{2010}]{lazio10}
{Lazio} T.~J.~W.,  {Shankland} P.~D.,  {Farrell} W.~M.,   {Blank} D.~L.,  2010,
  \mn@doi [\aj] {10.1088/0004-6256/140/6/1929}, \href
  {http://adsabs.harvard.edu/abs/2010AJ....140.1929L} {140, 1929}

\bibitem[\protect\citeauthoryear{{Lecavelier des Etangs}, {Sirothia},
  {Gopal-Krishna}  \& {Zarka}}{{Lecavelier des Etangs}
  et~al.}{2013}]{lecavelier13}
{Lecavelier des Etangs} A.,  {Sirothia} S.~K.,  {Gopal-Krishna}  {Zarka} P.,
  2013, \mn@doi [\aap] {10.1051/0004-6361/201219789}, \href
  {http://adsabs.harvard.edu/abs/2013A%26A...552A..65L} {552, A65}

\bibitem[\protect\citeauthoryear{{Mittag}, {Robrade}, {Schmitt}, {Hempelmann},
  {Gonz{\'a}lez-P{\'e}rez}  \& {Schr{\"o}der}}{{Mittag}
  et~al.}{2017}]{mittag17}
{Mittag} M.,  {Robrade} J.,  {Schmitt} J.~H.~M.~M.,  {Hempelmann} A.,
  {Gonz{\'a}lez-P{\'e}rez} J.~N.,   {Schr{\"o}der} K.~P.,  2017, \mn@doi [\aap]
  {10.1051/0004-6361/201629156}, \href
  {https://ui.adsabs.harvard.edu/abs/2017A&A...600A.119M} {600, A119}

\bibitem[\protect\citeauthoryear{{Nicholson} et~al.,}{{Nicholson}
  et~al.}{2016}]{nicholson16}
{Nicholson} B.~A.,  et~al., 2016, \mn@doi [\mnras] {10.1093/mnras/stw731},
  \href {https://ui.adsabs.harvard.edu/abs/2016MNRAS.459.1907N} {459, 1907}

\bibitem[\protect\citeauthoryear{{{\'O} Fionnag{\'a}in} et~al.,}{{{\'O}
  Fionnag{\'a}in} et~al.}{2019}]{ofionnagain19}
{{\'O} Fionnag{\'a}in} D.,  et~al., 2019, \mn@doi [\mnras]
  {10.1093/mnras/sty3132}, \href
  {https://ui.adsabs.harvard.edu/abs/2019MNRAS.483..873O} {483, 873}

\bibitem[\protect\citeauthoryear{{O'Gorman}, {Coughlan}, {Vlemmings},
  {Varenius}, {Sirothia}, {Ray}  \& {Olofsson}}{{O'Gorman}
  et~al.}{2018}]{ogorman18}
{O'Gorman} E.,  {Coughlan} C.~P.,  {Vlemmings} W.,  {Varenius} E.,  {Sirothia}
  S.,  {Ray} T.~P.,   {Olofsson} H.,  2018, \mn@doi [\aap]
  {10.1051/0004-6361/201731965}, \href
  {http://adsabs.harvard.edu/abs/2018A%26A...612A..52O} {612, A52}

\bibitem[\protect\citeauthoryear{{Panagia} \& {Felli}}{{Panagia} \&
  {Felli}}{1975}]{panagia75}
{Panagia} N.,  {Felli} M.,  1975, \aap, \href
  {https://ui.adsabs.harvard.edu/abs/1975A&A....39....1P} {39, 1}

\bibitem[\protect\citeauthoryear{{Parker}}{{Parker}}{1958}]{parker58}
{Parker} E.~N.,  1958, \mn@doi [\apj] {10.1086/146579}, \href
  {https://ui.adsabs.harvard.edu/abs/1958ApJ...128..664P} {128, 664}

\bibitem[\protect\citeauthoryear{{Polzin}, {Breton}, {Stappers},
  {Bhattacharyya}, {Janssen}, {Os{\l}owski}, {Roberts}  \& {Sobey}}{{Polzin}
  et~al.}{2019}]{polzin19}
{Polzin} E.~J.,  {Breton} R.~P.,  {Stappers} B.~W.,  {Bhattacharyya} B.,
  {Janssen} G.~H.,  {Os{\l}owski} S.,  {Roberts} M.~S.~E.,   {Sobey} C.,  2019,
  \mn@doi [\mnras] {10.1093/mnras/stz2579}, \href
  {https://ui.adsabs.harvard.edu/abs/2019MNRAS.490..889P} {490, 889}

\bibitem[\protect\citeauthoryear{{R{\'e}ville}, {Brun}, {Matt}, {Strugarek}  \&
  {Pinto}}{{R{\'e}ville} et~al.}{2015}]{reville15}
{R{\'e}ville} V.,  {Brun} A.~S.,  {Matt} S.~P.,  {Strugarek} A.,   {Pinto}
  R.~F.,  2015, \mn@doi [\apj] {10.1088/0004-637X/798/2/116}, \href
  {https://ui.adsabs.harvard.edu/abs/2015ApJ...798..116R} {798, 116}

\bibitem[\protect\citeauthoryear{{Roberts}}{{Roberts}}{2011}]{roberts11}
{Roberts} M. S.~E.,  2011, in {Burgay} M.,  {D'Amico} N.,  {Esposito} P.,
  {Pellizzoni} A.,   {Possenti} A.,  eds,  American Institute of Physics
  Conference Series Vol. 1357, American Institute of Physics Conference Series.
  pp 127--130 (\mn@eprint {arXiv} {1103.0819}), \mn@doi{10.1063/1.3615095}

\bibitem[\protect\citeauthoryear{{Rybicki} \& {Lightman}}{{Rybicki} \&
  {Lightman}}{1986}]{rybicki86}
{Rybicki} G.~B.,  {Lightman} A.~P.,  1986, {Radiative Processes in
  Astrophysics}

\bibitem[\protect\citeauthoryear{{Sirothia}, {Lecavelier des Etangs},
  {Gopal-Krishna}, {Kantharia}  \& {Ishwar-Chandra}}{{Sirothia}
  et~al.}{2014}]{sirothia14}
{Sirothia} S.~K.,  {Lecavelier des Etangs} A.,  {Gopal-Krishna} {Kantharia}
  N.~G.,   {Ishwar-Chandra} C.~H.,  2014, \mn@doi [\aap]
  {10.1051/0004-6361/201321571}, \href
  {http://adsabs.harvard.edu/abs/2014A%26A...562A.108S} {562, A108}

\bibitem[\protect\citeauthoryear{{Smith}, {Collier Cameron}, {Greaves},
  {Jardine}, {Langston}  \& {Backer}}{{Smith} et~al.}{2009}]{smith09}
{Smith} A.~M.~S.,  {Collier Cameron} A.,  {Greaves} J.,  {Jardine} M.,
  {Langston} G.,   {Backer} D.,  2009, \mn@doi [\mnras]
  {10.1111/j.1365-2966.2009.14510.x}, \href
  {http://adsabs.harvard.edu/abs/2009MNRAS.395..335S} {395, 335}

\bibitem[\protect\citeauthoryear{{Turner}, {Grie{\ss}meier}, {Zarka}  \&
  {Vasylieva}}{{Turner} et~al.}{2019}]{turner19}
{Turner} J.~D.,  {Grie{\ss}meier} J.-M.,  {Zarka} P.,   {Vasylieva} I.,  2019,
  \mn@doi [\aap] {10.1051/0004-6361/201832848}, \href
  {https://ui.adsabs.harvard.edu/abs/2019A&A...624A..40T} {624, A40}

\bibitem[\protect\citeauthoryear{{Vidotto} \& {Bourrier}}{{Vidotto} \&
  {Bourrier}}{2017}]{vidotto17b}
{Vidotto} A.~A.,  {Bourrier} V.,  2017, \mn@doi [\mnras]
  {10.1093/mnras/stx1543}, \href
  {https://ui.adsabs.harvard.edu/abs/2017MNRAS.470.4026V} {470, 4026}

\bibitem[\protect\citeauthoryear{{Vidotto} \& {Donati}}{{Vidotto} \&
  {Donati}}{2017}]{vidotto17}
{Vidotto} A.~A.,  {Donati} J.~F.,  2017, \mn@doi [\aap]
  {10.1051/0004-6361/201629700}, \href
  {https://ui.adsabs.harvard.edu/abs/2017A&A...602A..39V} {602, A39}

\bibitem[\protect\citeauthoryear{{Weber} \& {Davis}}{{Weber} \&
  {Davis}}{1967}]{weber67}
{Weber} E.~J.,  {Davis} Leverett J.,  1967, \mn@doi [\apj] {10.1086/149138},
  \href {https://ui.adsabs.harvard.edu/abs/1967ApJ...148..217W} {148, 217}

\bibitem[\protect\citeauthoryear{{Weber}, {Erkaev}, {Ivanov}, {Odert},
  {Grie{\ss}meier}, {Fossati}, {Lammer}  \& {Rucker}}{{Weber}
  et~al.}{2018}]{weber18}
{Weber} C.,  {Erkaev} N.~V.,  {Ivanov} V.~A.,  {Odert} P.,  {Grie{\ss}meier}
  J.~M.,  {Fossati} L.,  {Lammer} H.,   {Rucker} H.~O.,  2018, \mn@doi [\mnras]
  {10.1093/mnras/sty2079}, \href
  {https://ui.adsabs.harvard.edu/abs/2018MNRAS.480.3680W} {480, 3680}

\bibitem[\protect\citeauthoryear{{Wood}}{{Wood}}{2004}]{wood04}
{Wood} B.~E.,  2004, \mn@doi [Living Reviews in Solar Physics]
  {10.12942/lrsp-2004-2}, \href
  {http://adsabs.harvard.edu/abs/2004LRSP....1....2W} {1, 2}

\bibitem[\protect\citeauthoryear{{Wood}}{{Wood}}{2018}]{wood18}
{Wood} B.~E.,  2018, in Journal of Physics Conference Series. p. 012028
  (\mn@eprint {arXiv} {1809.01109}), \mn@doi{10.1088/1742-6596/1100/1/012028}

\bibitem[\protect\citeauthoryear{{Zarka}, {Treumann}, {Ryabov}  \&
  {Ryabov}}{{Zarka} et~al.}{2001}]{zarka01}
{Zarka} P.,  {Treumann} R.~A.,  {Ryabov} B.~P.,   {Ryabov} V.~B.,  2001,
  \mn@doi [\apss] {10.1023/A:1012221527425}, \href
  {https://ui.adsabs.harvard.edu/abs/2001Ap&SS.277..293Z} {277, 293}

\bibitem[\protect\citeauthoryear{{Zarka}, {Cecconi}  \& {Kurth}}{{Zarka}
  et~al.}{2004}]{zarka04}
{Zarka} P.,  {Cecconi} B.,   {Kurth} W.~S.,  2004, \mn@doi [Journal of
  Geophysical Research (Space Physics)] {10.1029/2003JA010260}, \href
  {https://ui.adsabs.harvard.edu/abs/2004JGRA..109.9S15Z} {109, A09S15}

\makeatother
\end{thebibliography}

% ########################################################################
% End of document
% ########################################################################

% Typesetting comment
\bsp
\label{lastpage}
\end{document}